\documentclass[prb,twocolumn,groupeaddress]{revtex4-1}

\usepackage{amsmath}
\usepackage{amssymb}
\usepackage{xspace}
\usepackage{graphicx}
\usepackage{grffile}
\usepackage{nicefrac}
\usepackage{array}
\usepackage{color}

\newcolumntype{L}[1]{>{\raggedright\let\newline\\\arraybackslash\hspace{0pt}}m{#1}}
\newcolumntype{C}[1]{>{\centering\let\newline\\\arraybackslash\hspace{0pt}}m{#1}}
\newcolumntype{R}[1]{>{\raggedleft\let\newline\\\arraybackslash\hspace{0pt}}m{#1}}

\begin{document}

\title{Emergent flat-band physics in ${d}^{9\ensuremath{-}\ensuremath{\delta}}$
  multilayer nickelates}
\author{Frank Lechermann}
\affiliation{Institut f\"ur Theoretische Physik III, Ruhr-Universit\"at Bochum,
  D-44780 Bochum, Germany}

\begin{abstract}
  Recent experiments have shown that the reduced multilayer rare-earth (RE)
  nickel oxides of form RE$_{p+1}$Ni$_p$O$_{2p+2}$ may belong to the novel family of
  superconducting lanthanide nickelates. Here, the correlated electronic structure of
  Pr$_{4}$Ni$_3$O$_{8}$ and Nd$_{6}$Ni$_5$O$_{12}$ is studied by means of an
  advanced realistic many-body framework. It is revealed that the low-energy physics
  of both systems is dominated by an interplay of Ni-$d_{x^2-y^2}$ and Ni-$d_{z^2}$
  degrees of freedom. Whilst the Ni-$d_{x^2-y^2}$ orbitals are always highly correlated
  near an (orbital-selective) Mott-insulating regime, the Ni-$d_{z^2}$ orbitals give
  rise to intriguing non-dispersive features.
  At low temperature, the Pr compound still displays QP-like Ni-$d_{x^2-y^2}$-derived
  states at the Fermi level, but the interacting fermiology of the Nd compound is
  outshined by an emergent Ni-$d_{z^2}$-controlling flat band. These findings
  translate well to the previous characterization of doped infinite-layer nickelates,
  and hence further make the case for a mechanism of unconventional superconductivity
  which is distinct from the one in high-$T_{\rm c}$ cuprates.
\end{abstract}

\maketitle

\section{Introduction}
The physics of layered nickelate compounds has recently (re)gained enormous
interest due to the discovery of superconductivity in Sr-doped thin films
of infinite-layer NdNiO$_2$~\cite{li19,zen20}. The rise of a novel family of
superconducting oxides has been confirmed by subsequent same findings in
akin systems of PrNiO$_2$~\cite{osa20-1,osa20-2} as well as of 
LaNiO$_2$ type~\cite{osa21,zen21}. 

In fact, the quest for superconductivity in $p$-layered nickelates 
RE$_{p+1}$Ni$_p$O$_{3p+1(,2p+2)}$ of Ruddlesden-Popper(-like) structure with 
rare-earth ion RE (the oxygen stoichiometry in braces marking the 
topotactically reduced compounds without apical O) has a rather long 
history~\cite{cre83,lac92,cho96,hay99,ani99,lee04,pol06,par10,zha16,bot16}.
It started off from investigating single-layer La$_2$NiO$_4$, which is structurally
most akin to the Cu-$d_{x^2-y^2}$-driven high-$T_{\rm c}$ cuprate La$_2$CuO$_4$. But
in that setting, the Ni$^{2+}(3d^8)$ ion turns out orbital-wise rather
different from the reference Cu$^{+}(3d^9)$ ion. Further layering and topotactic
reduction should therefore be most effective in singling out the Ni-$d_{x^2-y^2}$ 
orbital in a low-energy regime for some (effective) Ni$^+$ 
setting~\cite{lac92,pol06}. 
However, single crystals for the ``straightforward'' infinite-layer
$(p$$\rightarrow$$\infty)$ materials
were absent until very recently~\cite{pup21}, and therefore focus consolidated 
on the multilayer systems with small $p$. And indeed, single crystals of the
reduced $p=3$ system were successfully prepared by 
Zhang~{\sl et al.}~\cite{zha16}. Yet superconductivity 
was still not detected, but then eventually the thin-film realization of 
hole-doped infinite-layer nickelates made the breaktrough~\cite{li19,zen20}.
Interestingly, the optimal hole doping for the infinite-layer compounds
can effectively also be realized for systems with $p\sim 5,6$
layers~\cite{zha16,liz20,mit21}. Actually, in a very recent study
Pan {\sl et al.}~\cite{pan21} reported superconductivity in
the reduced $p=5$ compound Nd$_{6}$Ni$_5$O$_{12}$.

This story has a further level of complexity, since the actual electronic
structure and the resulting superconducting scenario of the infinite-layer
nickelates might be not of straightforward cuprate kind and is still 
heavily debated (see e.g. Refs.~\onlinecite{bot21,zhatao21,pic21,mit21,che22}
for recent reviews). Main issues are the
degree of correlation strength, the role of apparent self-doping
bands, and the relevance of a Ni-$d_{x^2-y^2}$ vs. a Ni-multiorbital setting
at low energy. Against this challenging background, we here report a
comparing first-principles many-body investigation of Pr$_{4}$Ni$_3$O$_{8}$ $(p=3)$
and the recently highlighted Nd$_{6}$Ni$_5$O$_{12}$ $(p=5)$. 

In our previous studies of infinite-layer 
nickelates~\cite{lec20-1,lec20-2,lec21-1,lec21-2}, we argued that besides local 
Coulomb interactions on Ni, an effective inclusion of such local interactions on the 
O sites is indispensable to arrive at a reliable picture of the correlated electronic 
structure. Based on that viewpoint, it turns out that the infinite-layer nickelate 
physics is dominated by the intriguing interplay between 
Ni-$e_g$ $\{d_{z^2},d_{x^2-y^2}\}$ multiorbital degrees of freedom. While Ni-$d_{x^2-y^2}$
is orbital-selective Mott insulating and hardly doped with holes, the 
Ni-$d_{z^2}$ orbital eagerly collects hole carriers and shifts 
as a flat band in the $k_z=1/2$ plane of the Brillouin zone across the Fermi
level in the superconducting doping region. Main result of the present work
is that this emergent flat band physics in front of a highly correlated
Ni-$d_{x^2-y^2}$ state is indeed also relevant for ${d}^{9\ensuremath{-}\ensuremath{\delta}}$
multilayer nickelates, consistent with the comparable effective hole doping for 
superconducting infinite-layer and multilayer nickelate. Importantly, for the
Pr $(p=3)$ compound, the flat band has already passed the Fermi level, dissolving
into incoherent (Hund-driven) excitations. However,
it is about ``in resonance'' with the Fermi energy for the Nd $(p=5$) compound.
This underlines the decisive role of the Ni-$d_{z^2}$-dominated flat-band physics
for the stabilization of the superconducting phase in infinite- and multilayer
nickelates.

\section{Theoretical approach\label{sec:med}}
To reveal the correlated electronic structure of Pr$_{4}$Ni$_3$O$_{8}$ and
Nd$_{6}$Ni$_5$O$_{12}$, the charge self-consistent combination~\cite{sav01,gri12} 
of density functional theory (DFT), dynamical mean-field theory (DMFT) and 
self-interaction correction (SIC) is employed~\cite{lec19}. The DFT part of this 
DFT+sicDMFT scheme builds up on a mixed-basis pseudopotential 
framework~\cite{els90,lec02,mbpp_code} in the local density approximation (LDA).
We address the Coulomb interactions on oxygen furthermore beyond DFT within SIC 
on the pseudopotential level~\cite{vog96,fil03,kor10}. Whereas the O$(2s)$ orbital 
is by default fully corrected with a weight factor $w_{2s}=1.0$, the reasonable 
choice~\cite{kor10,lec19} $w_{2p}=0.8$ is utilized for the O$(2p)$ orbitals. 
The further screening parameter $\alpha$ for this SIC pseudoptential is chosen 
also as $\alpha=0.8$, such that the SIC inclusion on O asks for one additional 
parameter in the overall computational scheme. For a further discussion of the
relevance of the use of SIC on O in layered nickelates we refer to 
Ref.~\onlinecite{lec21-1}. Finally, the Ni sites act as quantum impurity problems in
multi-site realistic DMFT, with the site-resolved correlated subspace consisting
of the full Ni$(3d)$ shell, respectively. The whole scheme is converged until
self-consistency in the charge density and the self-energies is reached. Note again
that the SIC incorporation enters the given framework as a modified 
(pseudo)potential, and is not a 'fixed correction' or 'shift' on the sole DFT 
level.

In the following the calculational settings are described in more detail.
A $11\times 11\times 11$ k-point mesh is utilized for Pr$_{4}$Ni$_3$O$_{8}$,
and a $5\times 5\times 5$ one for Nd$_{6}$Ni$_5$O$_{12}$. The plane-wave cutoff 
energy is set to $E_{\rm cut}=13$\,Ry and localized basis orbitals are introduced for
Pr/Nd$(5d)$, Ni$(3d)$ as well as O$(2s,2p)$.
The Pr/Nd$(4f)$ states are put in the pseudopotential frozen core, since they are
not decisive for the key physics of infinite-layer nickelates~\cite{zha20}. 
The $j$-resolved frozen occupation of the Pr/Nd$(4f)$ shell is chosen with small 
resulting moment and only scalar-relativistic effects enter the general 
pseudopotential generation. The role of spin-orbit effects in the overall crystal 
calculations is neglected. The DMFT correlated subspace on each Ni site is governed 
by a full Slater Hamiltonian applied to the Ni$(3d)$ projected-local 
orbitals~\cite{ama08}. 
The projection is performed on the 
$N({\rm O})\times 3+N({\rm Ni})\times 5+N({\rm RE})$ Kohn-Sham (KS) 
states above the dominant O$(2s)$ bands, associated with O$(2p)$, Ni$(3d)$ and 
possibly relevant RE-based bands. Here, $N$(element) amounts to the element-specific
number of atoms in the unit cell, e.g. four for Pr$_{4}$Ni$_3$O$_{8}$ and 
five for Nd$_{6}$Ni$_5$O$_{12}$. This choice resembles the use of one addititonal KS 
state, i.e. $6+5+1$ in infinite-layer nickelates 
RENiO$_2$~\cite{lec20-1,lec20-2,lec21-1,lec21-2}.
A Hubbard $U=10$\,eV and a Hund exchange $J_{\rm H}=1$\,eV prove reasonable for 
this choice of the energy window~\cite{lec20-1,lec19}. The fully-localized-limit 
double-counting scheme~\cite{ani93} is applied. Continuous-time quantum Monte Carlo 
in hybridzation expansion~\cite{wer06} as implemented in the TRIQS 
code~\cite{par15,set16} is used to solve the DMFT problem. Two different 
system temperatures, namely $T=193$\,K and $T=50$\,K are chosen to reveal relevant
coherence effects. Up to $1.5\cdot 10^{9}$ Monte-Carlo sweeps are performed 
to reach convergence. A Matsubara mesh of 1025(2049) frequencies is
used to account for the higher(lower)-temperature regime. 
Maximum-entropy~\cite{jar96} and Pad{\'e}~\cite{vid77} methods are employed for 
the analytical continuation from Matsubara space onto the real-frequency 
axis. All calculations are conducted for a paramagnetic regime, respectively.

\section{Initial characterization of the $p=3,5$-layer nickelates}
\begin{figure*}[t]
\includegraphics*[width=17.75cm]{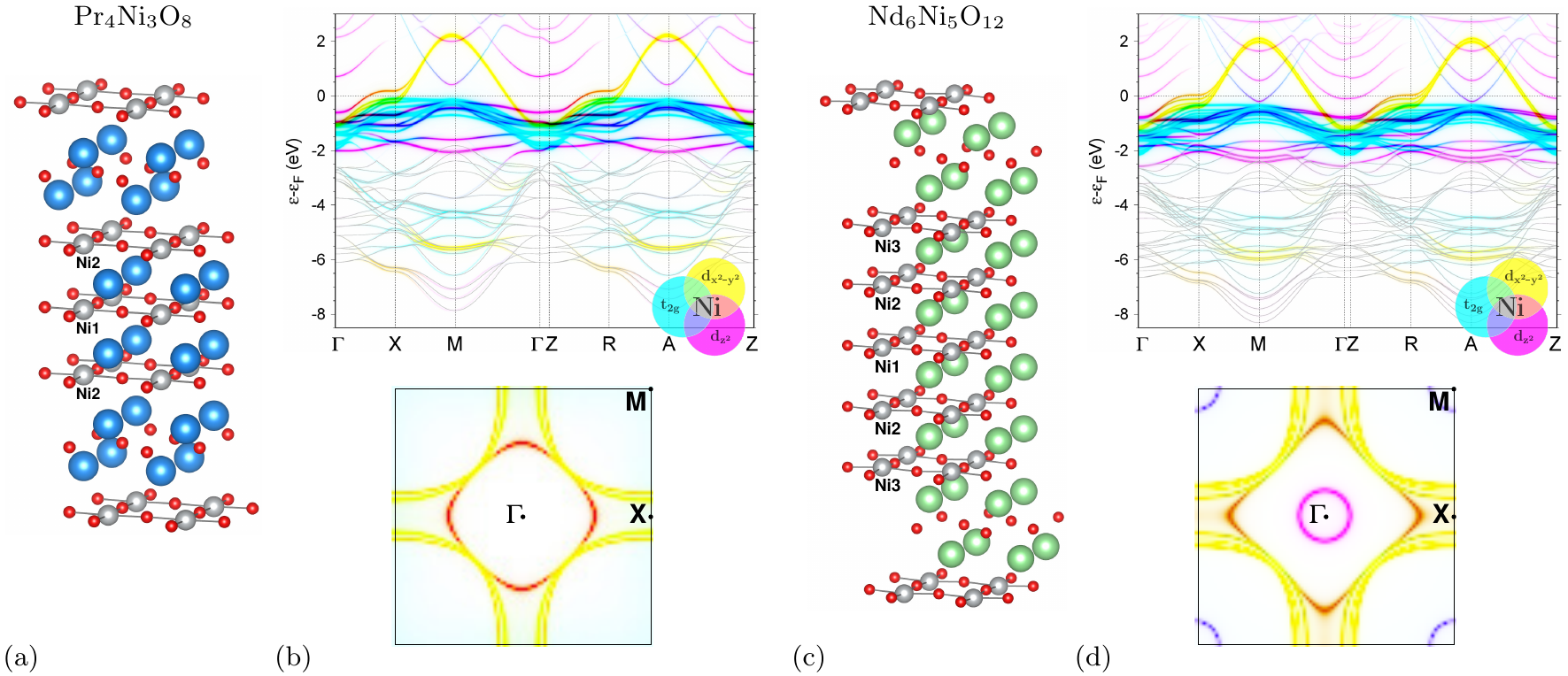}
\caption{(color online) Crystal structure and DFT spectrum of Pr$_{4}$Ni$_3$O$_{8}$ 
  (a,b) and Nd$_{6}$Ni$_5$O$_{12}$ (c,d). (a) Structure of the $p=3$ Pr compound with
  Pr (blue), Ni(grey) and oxygen (red); Ni1 and Ni2 class with associated NiO$_2$
  layers are indicated. (b) Band structure along high-symmetry lines (top) and
  Fermi surface (bottom) of
  Pr$_{4}$Ni$_3$O$_{8}$ in Ni$(3d)$-fatspec (see text) representation. Note that
  mixed colors represent joint contributions from the involved orbitals, e.g.
  red color corresponds to a mixing of Ni-$d_{x^2-y^2}$ (yellow) and Ni-$d_{z^2}$
  (magenta). The O$(2p)$-based block of bands is drawn with full grey lines.
  (c) Structure of the $p=5$ Nd compound with Nd (green), Ni(grey) and oxygen (red);
  Ni1, Ni2 and Ni3 class with associated NiO$_2$ layers are again indicated.
  (d) Same as (b), but for Nd$_{6}$Ni$_5$O$_{12}$.
}
\label{fig:dft}
\end{figure*}

The reduced layered nickelates of type RE$_{p+1}$Ni$_p$O$_{2p+2}$
(see Figs.~\ref{fig:dft}a,c) crystallize in a tetragonal $I4/mmm$ space group and host
blocks of $p$ NiO$_2$ square-lattice layers separated by RE layers. These blocks of
NiO$_2$ and RE layers are again separated by fluorite-like REO slabs, increasing the
two-dimensional character of the compounds. The here studied systems
Pr$_{4}$Ni$_3$O$_{8}$ and Nd$_{6}$Ni$_5$O$_{12}$ are chosen for the following
reasons. First, the nominal average Ni$(3d)$ occupation amounts to~\cite{lab21}
$d^{8.67}$ for $p=3$ and
to $d^{8.80}$ for $p=5$. Since the superconducting region of the infinite-layer nickelates 
nominally spans between $\sim d^{8.75-8.85}$, the $p=5$ compound should effectively be prone
to superconductivity, while the $p=3$ compound should reside in the hole-overdoped regime.
Thin films of Nd$_{6}$Ni$_5$O$_{12}$ have indeed been shown to exhibit superconducting
properties~\cite{pan21}.
For the $p=3$ case we select Pr$_{4}$Ni$_3$O$_{8}$ since it was prepared in
single-crystal form~\cite{zha17-2,linjq21} with metallic behavior down to lowest
temperatures.

The trilayer Pr compound has two symmetry-inequivalent Ni classes: the Ni1 class
corresponding
with the inner NiO$_2$ layer and the Ni2 class associated with the outer Ni2 layer.
The quintuple-layer Nd compound has three such different classes:
Ni1 for inner, Ni2 for intermediate
and Ni3 for outer NiO$_2$ layer. Lattice parameters $a=3.935$\,\AA\,and 
$c=25.485$\,\AA\,, as well as atomic positions are here overtaken from 
experiment~\cite{zha17-2} for Pr$_{4}$Ni$_3$O$_{8}$. In the case of 
Nd$_{6}$Ni$_5$O$_{12}$, the experimental $c$-axis
parameter of the thin-film study~\cite{pan21}, i.e. $c=38.8$\,\AA, and the
inplane $a=3.92$\,\AA\, from NdNiO$_2$~\cite{li19} are used. The atomic positions for
the Nd compound are obtained via DFT structural relaxation using the
generalized-gradient approximation. This results in a somewhat stronger buckling of the
non-Ni1O$_2$ layers compared to the one in the Pr compound
(compare Figs.~\ref{fig:dft}a,c).

The DFT characterization (to be compared with previous
studies~\cite{nic20,kar20-2,lab21,pan21,wor21,lab21-2}) of both multilayer compounds is 
summarized in Figs.~\ref{fig:dft}b,d. We choose a {\sl fatspec} representation for the
band structures. It colors the spectrum according to the orbital weight in the
spectral function $A({\bf k},\omega)$ at given $({\bf k},\omega)$ point. Of course, in
the DFT limit, $A({\bf k},\omega)$ reduces to the KS band structure. Note that this
fatspec representation differs slightly from the usual 'fatband' picture, since
overlayed/crossing bands and thus non-hybridized energy areas may also appear
as 'mixed-orbital' character. For instance, a Ni-$d_{z^2}$-dominated dispersion
(magenta-colored) starts in the top panel of Fig.~\ref{fig:dft}b at $\sim -1$\,eV and
closes at X with energy $\sim -0.75$\,eV. It appears  blue-colored inbetween,
because it overlays with Ni-$t_{2g}$ (cyan-colored) in that energy range. Surely,
the truly mixed hybridization on an individual dispersion is also correctly signalled.
Here, the fatspec representation focusses on the Ni$(3d)$ nature of the electronic
structure.

Starting with Pr$_{4}$Ni$_3$O$_{8}$, it is
seen that the Ni-$t_{2g}$-dominated bands are mostly full and only the Ni-$e_g$-derived
bands are partially occupied. Note that there is no self-doping band as in infinite-layer
nickelates~\cite{lec20-1}, the corresponding bands with sizable Ni-$d_{z^2}$, Ni-$t_{2g}$
and Pr$(5d)$ character are well above the Fermi level $\varepsilon_{\rm F}$. Along
$\Gamma$-X, one band of mixed Ni-$e_g$ type crosses $\varepsilon_{\rm F}$, whilst along
X-M there are two dominant Ni-$d_{x^2-y^2}$ dispersions at low-energy.
In other words, there are $p$ Ni-$e_g$ bands crossing the Fermi level
This results in a three-sheeted Fermi surface (FS) for the Pr compound, with
an electron-like mixed Ni-$e_g$ inner sheet centred around $\Gamma$ and two close-running
Ni-$d_{x^2-y^2}$-based hole-like sheets. The latter are reminiscent of the hallmark hole
sheet in high-$T_{\rm c}$ cuprates.

The DFT spectrum of the Nd compound displays further-increased complexity (see
Fig.~\ref{fig:dft}d). Now five
Ni-$e_g$-based bands cross $\varepsilon_{\rm F}$; a strongly-mixed one again
along $\Gamma$-X and four along X-M. Three out of these latter four bands are
dominantly of Ni-$d_{x^2-y^2}$ kind and the highest-lying one with some Ni-$d_{z^2}$
mixed in. Note that the mixed Ni-$e_g$ dispersion crossing in $\Gamma$-X direction
has its van-Hove point at X closer to the Fermi level than in the Pr compound.
Furthermore, there are two additional self-doping bands, one close to $\Gamma$
and the second close to M. The first one has sizable Ni-$d_{z^2}$ flavor, whereas
the one next to M has also Ni-$t_{2g}$ mixed in. Accordingly, besides the five-sheeted
dominant Ni-$e_g$ fermiology, the FS exhibits two additional electron-like pockets
around $\Gamma$ and M. As a further difference to Pr$_{4}$Ni$_3$O$_{8}$, the most-mixed
Ni-$e_g$ sheet (i.e. red-colored in bottom panel of Figs.\ref{fig:dft}b,d) is larger
and has more square shape in the Nd compound. Interestingly, the self-doping bands are
seemingly missing in the $p=5$ compound of La type~\cite{lab21,lab21-2}, but are also
observed in the DFT result for Nd$_{6}$Ni$_5$O$_{12}$ from Ref.~\onlinecite{wor21}.

Finally, we want to comment on the charge-transfer energy
$\Delta=\varepsilon_d-\varepsilon_p$, whereby $\varepsilon_{d,p}$ are the respective
band centre of Ni$(3d)$ and O$(2p)$. Using the DFT+sic approach, a value
$\Delta_{\rm NdNiO_2}=5.0$\,eV was obtained for infinite-layer NdNiO$_2$~\cite{lec20-1}.
For the present multilayer systems, calculations from the same perspective result
in $\Delta_{\rm Nd_6Ni_5O_{12}}=4.2$\,eV and $\Delta_{\rm Pr_3Ni_3O_8}=3.9$\,eV.
This drop of the charge-transfer energy with an increase of the Ni effective oxidation 
state is in line with previous findings~\cite{boc92,lec21-2}.

\section{DFT+\MakeLowercase{sic}DMFT results}
\subsection{P\MakeLowercase{r}$_{4}$N\MakeLowercase{i}$_3$O$_{8}$}
\begin{figure*}[t]
\includegraphics*[width=17.75cm]{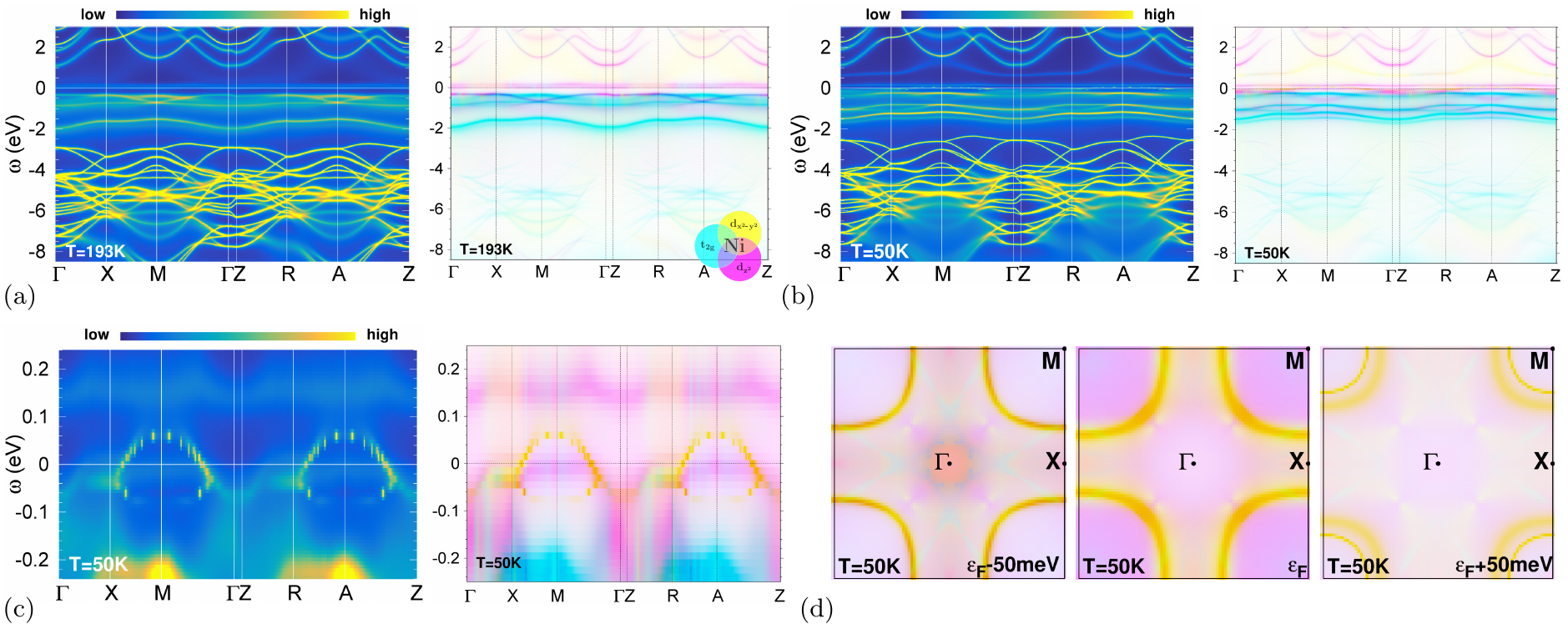}
\caption{(color online) {\bf k}-resolved DFT+sicDMFT spectrum Pr$_{4}$Ni$_3$O$_{8}$.
  (a) Complete $A({\bf k},\omega)$ along high-symmetry lines (left) and Ni$(3d)$ fatspec
  representation (right), both at $T=193$\,K. (b) Same as (a) but at $T=50$\,K.
  (c) Same as (b) but focussing on low-energy window around $\varepsilon_{\rm F}$.
  (d) $k_z=0$ constant-energy surfaces at $T=50$\,K, from left to right: 
  $\varepsilon_{\rm F}$-50 meV, $\varepsilon_{\rm F}$, $\varepsilon_{\rm F}$+50 meV.}
\label{fig:k438}
\end{figure*}
\begin{figure}[t]
\includegraphics*[width=8.5cm]{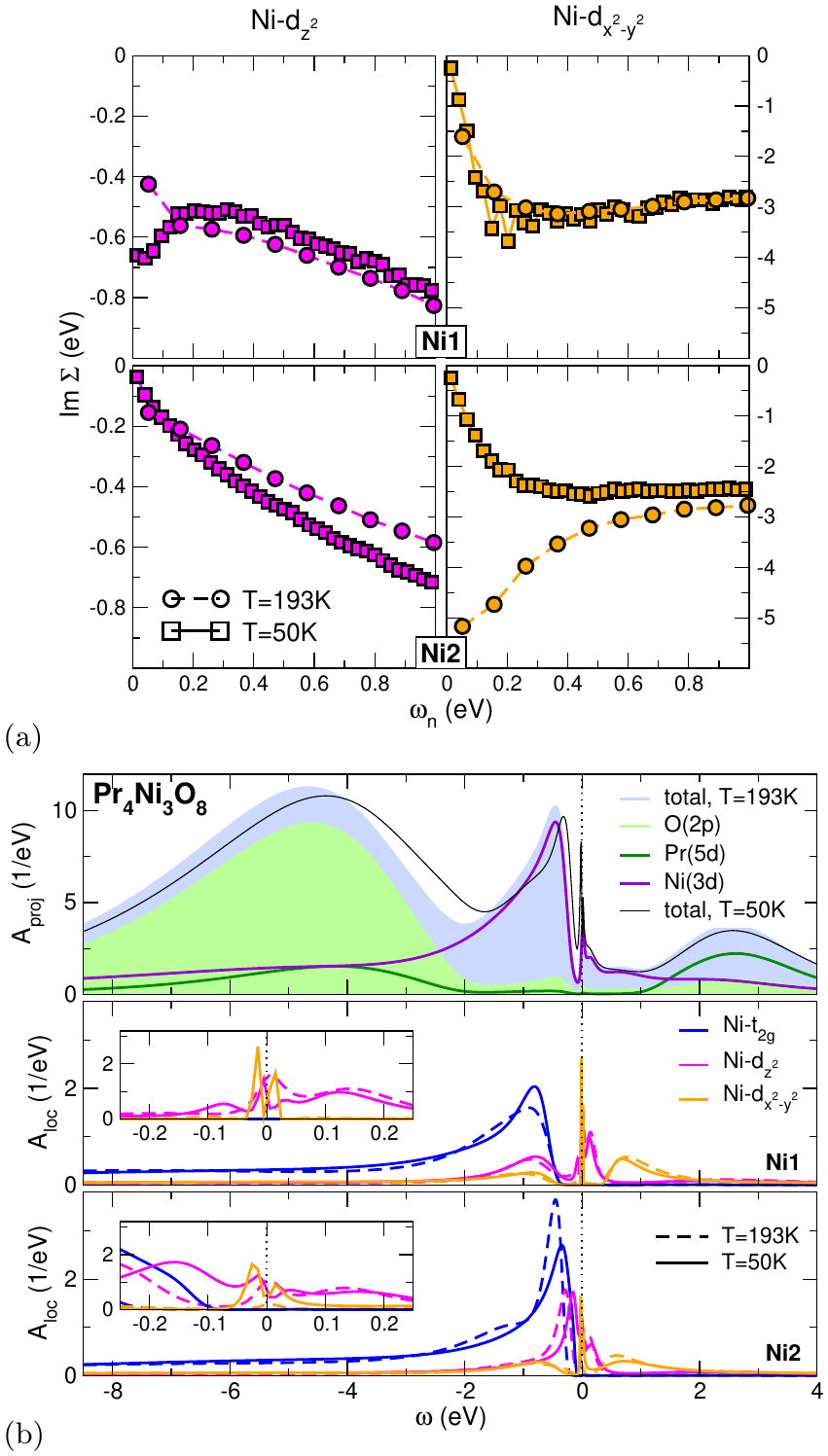}
\caption{(color online) Ni-$e_g$ self-energies (a) and
  {\bf k}-integrated DFT+sicDMFT spectrum (b) of Pr$_{4}$Ni$_3$O$_{8}$.
(a) Imaginary part of the  Ni-$d_{z^2}$ (left) and the Ni-$d_{x^2-y^2}$ (right)
Matsubara self-energy $\Sigma(i\omega_{\rm n})$ for Ni1 (top) and Ni2 (bottom)
at $T=193$\,K (circle-dashed) and $T=50$\,K (square-full).
(b) Top: total as well as element- and site-resolved projected spectral function;
middle and bottom: local Ni1$(3d)$ and Ni2$(3d)$ spectral function at $T=193$\,K
(dashed) and $T=50$\,K (full), respectively. Insets: low-energy window.}
\label{fig:totself438}
\end{figure}

Let us turn to the interacting problem beyond DFT by first focussing on the trilayer
Pr compound. Intuitively, from the anticipated similarity to the overdoped infinite-layer
system~\cite{lec21-1}, one may expects the correlated electronic structure of 
Pr$_{4}$Ni$_3$O$_{8}$ to be in a Hund(-like) regime where coherence effects and 
possible non-Fermi-liquid (NFL) characteristics due to an interplay of Ni-$d_{z^2}$
and Ni-$d_{x^2-y^2}$ may be important.
Figure~\ref{fig:k438}a shows the {\bf k}-resolved spectrum at $T=193$\,K 
($\beta=1/T=60$\,eV$^{-1}$), with an indeed absence of well-defined quasiparticle (QP)
dispersions at the Fermi level. The Ni-$d_{x^2-y^2}$-dominated states are very 
strongly correlated with large scattering rate for Ni1 and an even diverging imaginary
Matsubara self-energy for Ni2 (cf. orange circle-dashed lines in right panel of 
Fig.~\ref{fig:totself438}a).
The fatspec representation reveals that the Ni-$t_{2g}$-dominated bands remain well below
$\varepsilon_{\rm F}$. Furthermore, it displays that the near-dispersionless spectral
weight at $\sim$0.15\,eV above the Fermi level is of Ni-$d_{z^2}$ character. Further such
character with somewhat more (incoherent) dispersion may be observed in a similar energy
range below $\varepsilon_{\rm F}$. 

Lowering the temperature to $T=50$\,K leads to a significant increase of the 
Ni-$d_{x^2-y^2}$ coherence, resulting in highly-renormalized QP dispersions as shown in
Fig.~\ref{fig:k438}b,c. This Fermi-liquid (FL) regime is also observable from the 
associated Ni-$d_{x^2-y^2}$ self-energies for both Ni sites (cf. orange square-full lines 
in right panel of Fig.~\ref{fig:totself438}a). The effective mass $m^*/m_{\rm DFT}\sim 14$
turns out very large for this orbital sector. 
In addition, notably, the impact of temperature, leading 
also to energy shifts for the Ni-$t_{2g}$-dominated bands, is stronger than in
stoichiometric infinite-layer NdNiO$_2$, where Ni-$d_{x^2-y^2}$ resides in an effective
Mott-insulating regime throughout the accessible temperature
range~\cite{lec20-1,lec20-2}. However as seen in Fig.~\ref{fig:k438}c, there is still
strong incoherent Ni-$d_{z^2}$ spectral weight at and close to the Fermi level, which
adds up to an appreciable total low-energy weight upon ${\bf k}$-integration for both
symmetry-inequivalent Ni sites (cf. Fig.~\ref{fig:totself438}b).
The incoherent nature of this Ni-$d_{z^2}$ weight is also
drawn from the displayed imaginary part of the corresponding self-energies in 
Fig.~\ref{fig:totself438}a (cf. magenta square-full lines in left panel): for Ni1 it shows
a clear NFL upturn at small Matsubara frequencies, and for Ni2 the bending in the same
frequency range is too strong to account for the linear-frequency behavior of a FL. In
other words, the electronic regime that one encounters in Pr$_{4}$Ni$_3$O$_{8}$ is that
of an partly incoherent metal with still highly-renormalized Ni-$d_{x^2-y^2}$ QP(-like)
dispersions. Concerning the aforementioned connection of effective Ni charge states,
this finding matches with a principal Hund-metal identification of the overdoped 
infinite-layer regime~\cite{lec21-1}.

The total spectral function $A(\omega)$ (see top panel of Fig.~\ref{fig:totself438}b),
besides the metallic weight at $\varepsilon_{\rm F}$, shows a dominant Ni$(3d)$ peak
at $\sim -0.5$\,eV and a dominant broader O$(2p)$ peak at $\sim-4.5$\,eV. Those dominant
$d$ and $p$ peaks are shifted towards the Fermi level compared to stoichiometric
NdNiO$_2$, again understandable from the doping connection. Note that there
is also some non-minor Pr$(5d)$ weight located within the broader O$(2p)$-dominated
peak, associated with the Pr sites in the fluorite-like REO slab. The
Zhang-Rice character\cite{zha88} of the low-energy spectral weight, i.e. a
spectral sharing between Ni$(3d)$ and O$(2p)$, is small; the latter weight
is dominantly of Ni$(3d)$ type. In view
of the Ni1,Ni2 differentiation, the inner Ni1 site exhibits a somewhat stronger
Ni-$d_{x^2-y^2}$ low-energy peak structure than the outer Ni2 site. But this weight
is of pseudogap structure for both sites, possibly due to the
coupling to the NFL-behaving Ni-$d_{z^2}$ states.

Table~\ref{tab:fill} provides the Ni$(3d)$ fillings, and while the values for
Ni-$d_{x^2-y^2}$ are always close to half filling, the Ni-$d_{z^2}$ orbitals are mainly
controlling the hole content. For Pr$_{4}$Ni$_3$O$_{8}$, the filling $n(d_{z^2})$
is lowest with an average value $\sim$\,1.45 and some larger filling for the outer
Ni2 site (translating to a slight overall larger $n(3d)$ filling for Ni2). Albeit
far off from a formal Ni1$^{2+}$,  Ni2$^{+}$ dichotomy~\cite{upt20},
the larger Ni$(3d)$ charge on Ni2 is qualitatively in line with this simple picturing.
Note that the theoretical filling values from DFT+sicDMFT, as usual depending on
e.g. the given definition of the correlated subspace, are generally somewhat smaller
than from the simplistic aforementioned oxidation-state analysis for $p=3,5$. From a
direct theory comparison to calculations for hole-overdoped
Nd$_{1-x}$Sr$_{x}$NiO$_2$ with $x=0.3$, the present $n(d_{z^2})$ for the
trilayer compound is also significantly lower. Thus the comparison
between multilayer and doped-infinite-layer physics has still to be performed with
caution.

Let us finally compare to experiment. Zhang {\sl et al.}~\cite{zha17-2} reported metallic
behavior down to lowest $T$, with however a rather high room-temperature resistivity of
67.2\,$\Omega$\,cm. Additionally, the specific-heat measurements from that work give rise
to a large $\gamma\sim$\,75 mJ\,mol$^{-1}$\,K$^{-2}$. In comparison,
for La$_{4}$Ni$_3$O$_{10}$ a value $\gamma\sim$\,15 mJ\,mol$^{-1}$\,K$^{-2}$
is obtained~\cite{rou20} and a mass
enhancement of about 3-4 infered. Disregarding possible Pr$(4f)$ physics, an enormous
formal mass enhancement of $\sim$\,15 for Pr$_{4}$Ni$_3$O$_{8}$ is thus not completely
out of range. Magnetic properties have been measured by
Huangfu~{\sl et al.}~\cite{hua20} showing complex magnetic behavior without ordering
down to lowest $T$. But the data hints to a coexistence of localized spins and itinerant
degrees of freedom.
\begin{table}[t]
\begin{ruledtabular}
\begin{tabular}{l|l|cc|c}
compound                & Ni site & $n(d_{z^2})$ & $n(d_{x^2-y^2})$ & $n(3d)$ \\[0.05cm] \hline
Pr$_{4}$Ni$_3$O$_{8}$   & Ni1     & 1.40         & 1.13             & 8.47 \\
                        & Ni2     & 1.48         & 1.11             & 8.53 \\[0.05cm] \hline
Nd$_{6}$Ni$_5$O$_{12}$  & Ni1     & 1.65         & 1.08             & 8.67 \\
                        & Ni2     & 1.54         & 1.09             & 8.57 \\
                        & Ni3     & 1.56         & 1.08             & 8.58 \\[0.05cm] \hline
NdNiO$_2$\hfill[\onlinecite{lec21-1}]                & Ni & 1.85  & 1.07  & 8.86 \\
Nd$_{0.85}$Sr$_{0.15}$NiO$_2$~[\onlinecite{lec21-1}] & Ni & 1.76  & 1.04  & 8.74 \\
Nd$_{0.70}$Sr$_{0.30}$NiO$_2$~[\onlinecite{lec21-1}] & Ni & 1.60  & 1.08  & 8.58 \\[0.05cm] \hline
\end{tabular}
\end{ruledtabular}
\caption{Site and orbital-resolved Ni$(3d)$ fillings $n$ in the $p=3,5$-layer compounds from DFT+sicDMFT
at $T=50$\,K. The Ni-$t_{2g}$ filling varies only marginally and is always close to $n(t_{2g})=5.94$.
The comparing data~\cite{lec21-1} for stoichiometric and hole-doped infinite-layer NdNiO$_2$ 
was obtained for $T=30$\,K.}
\label{tab:fill}
\end{table}

\subsection{N\MakeLowercase{d}$_{6}$N\MakeLowercase{i}$_5$O$_{12}$}
\begin{figure*}[t]
\includegraphics*[width=17.75cm]{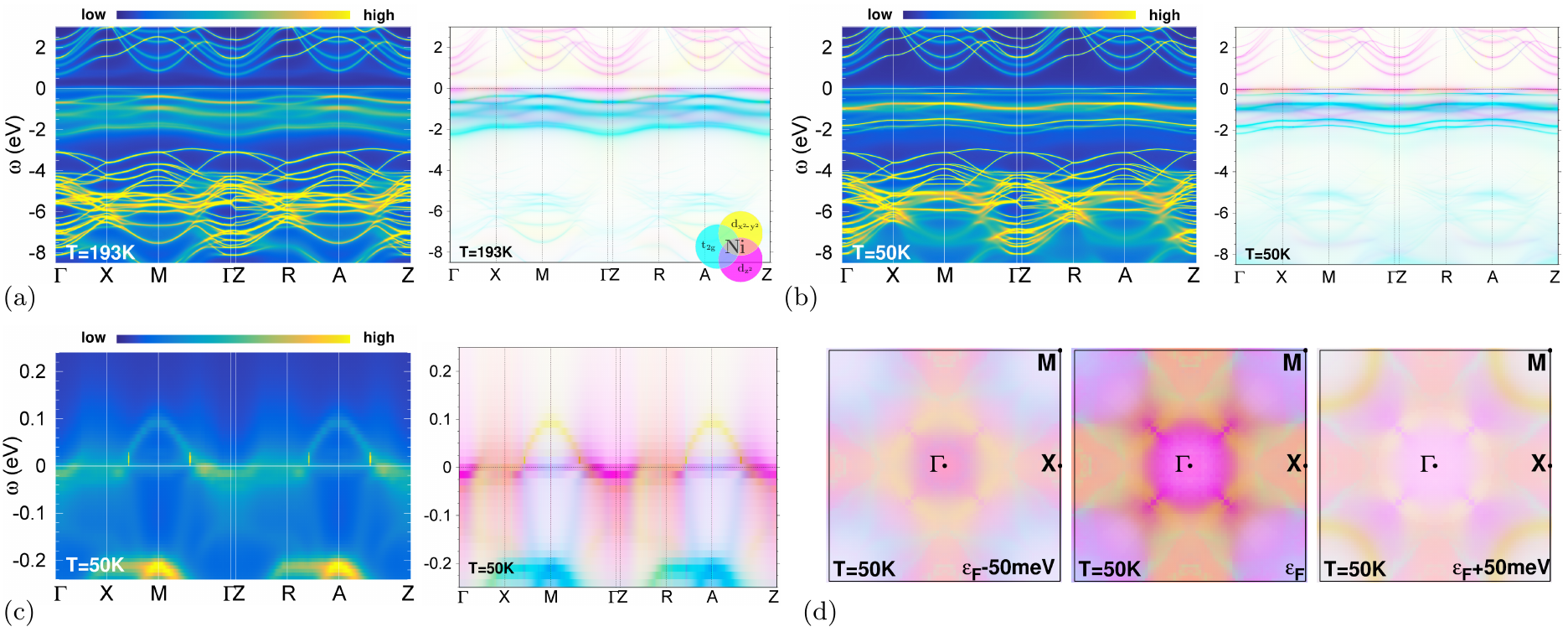}
\caption{(color online) Same as Fig.~\ref{fig:k438} but for Nd$_{6}$Ni$_5$O$_{12}$.}
\label{fig:k6512}
\end{figure*}
\begin{figure}[t]
\includegraphics*[width=8.5cm]{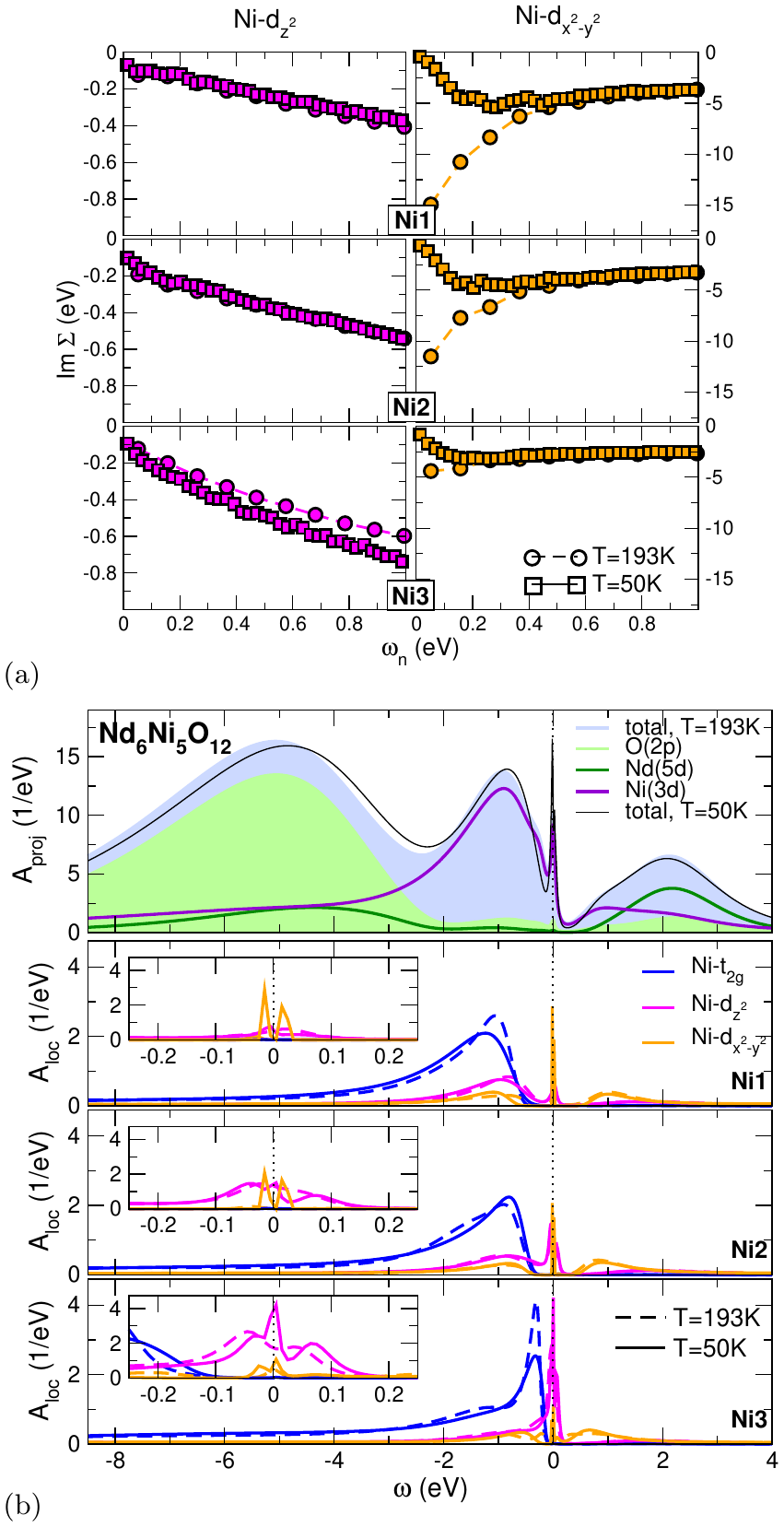}
\caption{(color online) Same as Fig.~\ref{fig:totself438} but for Nd$_{6}$Ni$_5$O$_{12}$.
  The Ni site-resolved data spans here over Ni1, Ni2 and Ni3.}
\label{fig:totself6512}
\end{figure}

Very much in parallel to the discussion of the correlated electronic structure of 
Pr$_{4}$Ni$_3$O$_{8}$ in the last section, the focus is henceforth on the $p=5$ Nd
compound. Figure.~\ref{fig:k6512} displays the {\bf k}-resolved data, with
obvious low-energy differences to the trilayer Pr system. The seemingly non-dispersive
spectral weight just above the Fermi level for the latter compound, appears here more
or less right at $\varepsilon_{\rm F}$ for both investigated temperatures
(cf. Fig.~\ref{fig:k6512}a,b). At $T=193$\,K, this Fermi-level spectral weight is
again very incoherent, and the Ni-$d_{x^2-y^2}$ self-energies are (about to) diverging
at low frequency (see Fig.~\ref{fig:totself6512}a). In the low-energ window of
Fig.~\ref{fig:k6512}c for $T=50$\,K, there is some strongly renormalized Ni-$d_{x^2-y^2}$
dispersion identifiable, but its intensity seems reduced compared to the
one in Pr$_{4}$Ni$_3$O$_{8}$.
Instead, the non-dispersive Ni-$d_{z^2}$ weight right at $\varepsilon_{\rm F}$ turns out
stronger for the present quintuple-layer Nd compound.

Note that the self-doping bands forming electron pockets around
$\Gamma$ and M in the previous DFT picture are shifted well-above the Fermi level and do
not play a role in the fermiology. Compared to infinite-layer NdNiO$_2$, where the
self-doping band 'survives' correlation effects and remains at the stoichiometric Fermi
level, this is understandable for two main reasons. First the charge-transfer from Pr$(5d)$
to Ni$(3d)$ in order to realize a favorable (near) half-filled Ni-$d_{x^2-y^2}$ scenario
in a strong-coupling limit is apparently larger for $p=5$. Second, the self-doping bands
carry quite some Ni-$d_{z^2}$ weight in both cases, however importantly, in the
quintuple-layer compound the filling $n(d_{z^2})$ is way smaller than in NdNiO$_2$
(cf. Tab.~\ref{tab:fill}). Therefore, correlations in Ni-$d_{z^2}$ are stronger, enabling a
larger correlation-induced shifting of the corresponding spectral weight. As for the
trilayer system, the Ni-$t_{2g}$-dominated states are mostly full. Yet there is a residual
Ni-$d_{z^2}$/$t_{2g}$ hybridization visible in the fatspec-representation panel of
Fig.~\ref{fig:k6512}c, extending towards the Fermi level.

Finally, the fermiology at $T=50$\,K looks quite different from the one of
Pr$_{4}$Ni$_3$O$_{8}$ and rather intriguing (see Fig.~\ref{fig:k6512}d) .
The role of the previously dominant Ni-$d_{x^2-y^2}$ states is hard to decypher, but much
weaker (to say the least). The original hole-like topology from this orbital sector becomes
only well tractable somewhat above the Fermi level close to the M point. Apparently, the
coherence scale for (possibly) robust QP-like Ni-$d_{x^2-y^2}$ excitations has not yet
been reached at $T=50$\,K for Nd$_{6}$Ni$_5$O$_{12}$. Also therefore, mass-enhancement
estimates are not too well-defined in this case. Yet from a crude examination of the
low-frequency Ni self-energy, the $m^*$ is about the twice the value than for the
trilayer Pr compound. The interacting Fermi surface appears to be
dominated by a correlated flat-band Ni-$d_{z^2}$ sheet around $\Gamma$.
Further conclusive FS details are hard to draw from the generally low-coherence
level at this temperature. Lowering the temperature much further is however numerically
tough for the given theoretical approach to these multilayer systems. But still, the
present results for the two $T$ scales render the fact robust, that emergent flat-band
physics of Ni-$d_{z^2}$ type rules the low-energy response. Form a site-averaged view
onto the Ni-$d_{z^2}$ self-energies shown in the left panel of Fig.~\ref{fig:totself6512}a,
the correlation effects on this flat band seem somewhat weaker than for the similarly
non-dispersive parts in the Pr compound. There are no very strong irregular features,
however the low-frequency bending for the Ni3 site is still NFL-like.

The {\bf k}-integrated spectra plotted in Fig.~\ref{fig:totself6512}b are for the
total content similar to the Pr$_{4}$Ni$_3$O$_{8}$ case, with a somewhat stronger
low-energy peak at $\varepsilon_{\rm F}$ and slightly deeper-energy location of the Ni$(3d)$
and O$(2p)$ peaks. The Zhang-Rice nature of the low-energy weight remains minor.
The site-resolved Ni spectra exhibit seemingly a monotonic Ni-$e_g$ evolution
at low energy, in the sense that the Ni-$d_{z^2}$ character grows from inner-to-outer
layer and vice versa for Ni-$d_{x^2-y^2}$ (with similar pseudogap structure as for the
trilayer compound). The emergent flat-band physics appears to be dominantly driven from
the outer Ni3 site. Interestingly, the Ni$(3d)$ charge hierachy between inner and outer
layers turn out reversed in Nd$_{6}$Ni$_5$O$_{12}$; the inner Ni1 site has a slightly
higher $d$ count than Ni2,Ni3 by about 0.1 electrons (see Tab.~\ref{tab:fill}).

Comparison to experiment is even more difficult in the quintuple-layer case. First,
there is so far only the thin-film work by Pan {\sl et al.}~\cite{pan21} on this
system. Second, even for $T=50$\,K, the theoretical spectrum/fermiology
has not yet completely settled in terms of coherence. The measured room-temperature
resistivity of $\sim 6$\,m$\Omega$\,cm yet is way smaller than for
single-crystal Pr$_{4}$Ni$_3$O$_{8}$, which could be in favor of a high density of
states at $\varepsilon_{\rm F}$. The experimental Hall coefficient is positive over the
whole accessible $T$ range. A negative sign of this coefficient for stoichiometric
NdNiO$_2$ is understandable from the self-doping electron pockets, and indeed, such
pockets are missing in the interacting regime of the present multilayer compounds.
But the very intriguing (incoherent) fermiology shown in Fig.~\ref{fig:k6512}d does
not really allow for more serious conclusions on this issue. At least,
many-body calculations~\cite{pet20} of the Hall
coefficient for optimally-doped infinite-layer NdNiO$_2$ results also in a positive
sign when having a previously-predicted~\cite{lec20-1} Ni-$d_{z^2}$ flat-band at the
Fermi level.

\section{Discussion}
The revealed data for the paramagnetic correlated electronic structure of
Pr$_{4}$Ni$_3$O$_{8}$ and Nd$_{6}$Ni$_5$O$_{12}$ shows that the two compounds share
certain similarities, but possess a rather different low-energy nature. Similar to both
systems is the way higher susceptibility to temperature effects than for the
stoichiometric NdNiO$_2$ compound. Akin to the infinite-layer nickelate is the
orbital-selective scenario of a highly correlated, (near) Mott-insulating
Ni-$d_{x^2-y^2}$ orbital sector
and an itinerant Ni-$d_{z^2}$ sector. However while the latter is close to complete
filling and only weakly-correlated in stoichiometric NdNiO$_2$, the correlation effects
are here more severe due to a lower filling with stronger overall Ni-$e_g$ inter-orbital
processes. These processes are apparently based on a sophisticated interplay between
orbital-selectivity, Hund-driven mechanisms and flat-band physics within the two-orbital
Ni-$e_g$ manifold. The very details of this interplay have to be addressed in a tailored
model-Hamiltonian study and are not subject of the present investigation. But as result
thereof, due to a different doping scenario, the trilayer Pr compound still displays
Ni-$d_{x^2-y^2}$ QP(-like) excitations within a background of non-dispersive incoherent
Ni-$d_{z^2}$ states, whereas Nd$_{6}$Ni$_5$O$_{12}$ exhibits a more incoherent Ni-$d_{x^2-y^2}$
dispersion coexisting with Ni-$d_{z^2}$ flat-band features.
\begin{figure}[b]
\includegraphics*[width=8.5cm]{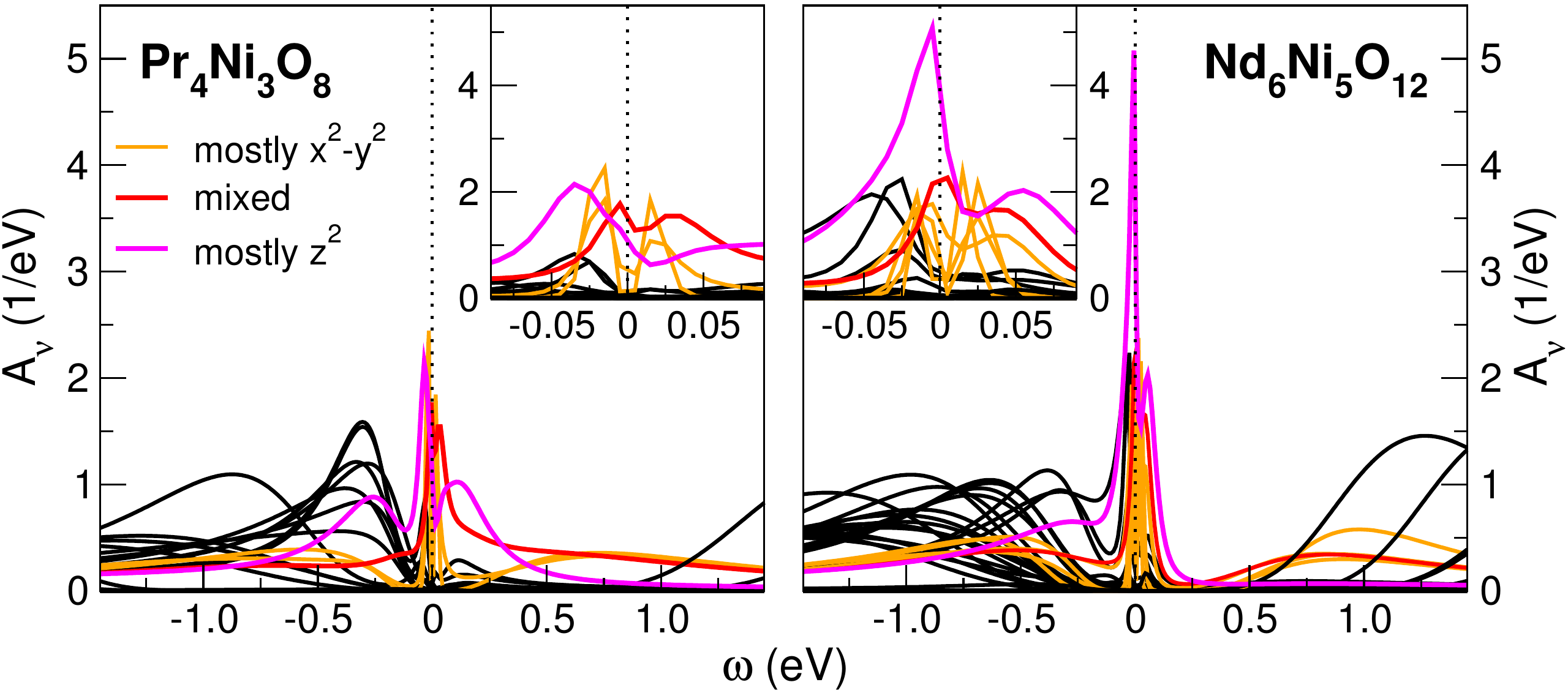}
\caption{(color online) Emergent flat-band character from the band-resolved correlated
  spectral function $A_\nu(\omega)$ (see text). Left: Pr$_{4}$Ni$_3$O$_{8}$ and right:
  Nd$_{6}$Ni$_5$O$_{12}$. Inset: focus around the Fermi lelve. The non-black colored
  functions denote the $\nu$-resolved $\varepsilon_{\rm F}$-crossing spectral
  weight with strongest near-$\omega=0$ contribution. Note that the coloring in terms
  of Ni-$d_{z^2}$ and Ni-$d_{x^2-y^2}$ marks only dominance and is for illustration,
  since this is not a fatspec/band plot.}
\label{fig:fb}
\end{figure}

Trying to rationalize this behavior, let us first focus on the Ni-$d_{x^2-y^2}$ part.
From DFT+sicDMFT calculations for NdNiO$_2$~\cite{lec20-1,lec20-2,lec21-1}, we learned
that Ni-$d_{x^2-y^2}$ is effectively Mott-localized at stoichiometry and remains incoherent
upon hole doping until a specific regime within the superconducting doping region. In the
infinite-layer overdoped region it becomes again incoherent. Thus the Ni-$d_{x^2-y^2}$
electrons behave very differently compared to the Zhang-Rice-bound
carriers in high-T$_{\rm c}$ cuprates,
where coherent QPs (modulo the notorious pseudogap regime) are robust over a wide
doping ranging starting just beyond stoichiometry. The strong charge-transfer character
of these cuprates, lacking in present nickelates, appears to be main origin for this.
But then it is still surprising that Pr$_{4}$Ni$_3$O$_{8}$ shows QP(-like) physics in
this nominally overdoped regime from the infinite-layer perspective. Reason could be
that a strong similarity between trilayer and infinite layer just breaks down in
the overdoped region. For instance, the additional degree of freedom of
Ni-site differentiation
might allow for electronic relaxation such as to render QP propagation more robust.
Another aspect may be the fact that the lanthanide neighboring compounds
La$_{4}$Ni$_3$O$_{8}$ and Nd$_{4}$Ni$_3$O$_{8}$ are experimentally identified insulating
at low temperature~\cite{zha16,qli21}
(though the trilayer Nd compound may be metallic in thin-film
geometry~\cite{pan21}). Thus the competition between metal and insulator is
seemingly tight, and a QP-based metallicity of the Pr compound may also be based on
subtleties that are outside the general rule.

Concerning the Ni-$d_{z^2}$ part, the difference between the 'non-dispersive' dispersion
just above the Fermi level in the trilayer Pr compound and the 'flat band' right at
$\varepsilon_{\rm F}$ in the quintuple-layer Nd compound has to be understood.
Key is the observation that as in doped NdNiO$_2$ a flat band (there in the
$k_z=1/2$ plane because of a missing layer differentiation) shifts through
the Fermi surface with growing hole content. This is illustrated
in Fig.~\ref{fig:fb}, where the band-resolved correlated spectral function
$A_\nu(\omega)$ is plotted. The function $A_\nu(\omega)$ results from upfolding the
local DMFT Green's function to the original crystal Hilbert space, i.e. displays
how strong each original KS band is dressed with correlations and contributes to
the total interacting spectrum. The red/magenta-highlighted $A_\nu$ mark the
strongest non-Mott-critical contributions to the Fermi spectral weight. In the case
of Pr$_{4}$Ni$_3$O$_{8}$ (left pane of Fig.~\ref{fig:fb}),
there are two such contributions, whereby especially the
magenta-colored one still shows sizable sign of correlation (i.e. sideband features).
The non-dispersive part at $\sim 0.15$\,eV above the Fermi level in Fig.~\ref{fig:k438}c
may be identified with the upper sideband of that latter $A_\nu$, peaking just a bit
below $\varepsilon_{\rm F}$. The sideband energy scale is too small for an ordinary upper
Hubbard band, so a Hund-mediated origin is likely. Note that hence only
two of the three original KS bands that cross the Fermi level are nearly Mott-insulating
(cf. inset, orange-colored lines in Fig.~\ref{fig:fb}). To make contact to the original
DFT band structure (see Fig~\ref{fig:dft}b): the here red-colored $A_\nu$ connects to
the highest Fermi-level KS band with mixed Ni-$d_{z^2}$/Ni-$d_{x^2-y^2}$ character; the
here magenta-colored $A_\nu$ connects to the highest-occupied KS Ni-$d_{z^2}$-dominated
dispersion, starting at $\sim -1$\,eV at $\Gamma$ and closing at $\sim -0.75$\,eV
at X. Hence intriguingly, a KS-occupied band is partly depleted in the interacting
regime and causes the non-dispersive sideband feature in DFT+sicDMFT. 

On the other hand for the Nd$_{6}$Ni$_5$O$_{12}$ compound, it is easily observed in
the right panel of Fig.~\ref{fig:fb} that there is one certain $A_\nu$ of strong
Ni-$d_{z^2}$ nature dominating right at the Fermi level. It is again connected to the
original-KS Ni-$d_{z^2}$-dominated dispersion just below the Ni-$d_{x^2-y^2}$-dominated
bands (cf. Fig~\ref{fig:dft}d), and describes the {\sl emergent flat band} at
$\varepsilon_{\rm F}$. Hence in both multilayer cases, the Ni-$d_{z^2}$ dispersions below
the Ni-$d_{x^2-y^2}$-dominated ones are an important key to the problem. However it
only evolves into an emergent flat band in the quintuple-layer case. In the trilayer
case, since at a nominal higher hole-doping level, the formerly flat-band part of
the Ni-$d_{z^2}$ dispersion has already crossed the Fermi level and dissolves
into stronger (Hund-)correlated states with incoherent spectral weight. Note that
as a further dfference, the topmost mixed Ni-$d_{z^2}$/Ni-$d_{x^2-y^2}$ dispersion
(red color in Fig.~\ref{fig:fb}) is stronger correlated in the Nd compound.

Let us finally note, that recent DFT+DMFT studies on multilayer
nickelates~\cite{kar20-2,wor21,lab21-2} differ from the here established low-energy
picture. From those works, Ni-$d_{z^2}$-influenced physics is mainly negligible and
moderately-to-strongly correlated Ni-$d_{x^2-y^2}$ physics (yet distant from an obvious
Mott-critical regime) is dominating. Key difference lies in the neglect of oxygen-based
correlations in Refs.~\onlinecite{kar20-2,wor21,lab21-2}, which renders the
layered nickelates
(much) weaker correlated than described here. Future experiments have to decide which
correlation regime and low-energy picture is more fitting. Note in that respect, that
a recent photoemssion study~\cite{chen21} for stoichiometric thin-films of
infinite-layer PrNiO$_2$ reports an electronic spectrum in very good accordance with
the DFT+sicDMFT predicted~\cite{lec20-1,lec21-1} spectrum for NdNiO$_2$.

\section{Conclusion}
The present DFT+sicDMFT study predicts emergent flat-band character of Ni-$d_{z^2}$
type in the quintuple-layer nickelate Nd$_{6}$Ni$_5$O$_{12}$. It is due to the formation
of near Mott-insulating Ni-$d_{x^2-y^2}$ states in an orbital-selective manner within
the Ni-$e_g$ subshell, in conjuction with the effective (optimal) hole doping compared
to stoichiometric infinite-layer NdNiO$_2$. Multiorbital correlations of seemingly
Hund-driven kind are also observable, but those are stronger for the trilayer nickelate
Pr$_{4}$Ni$_3$O$_{8}$. The latter compound is in a nominal hole-overdoped regime compared
to NdNiO$_2$, however still displays as QP-like Ni-$d_{x^2-y^2}$-dominated fermiology.
Whilst the principal mapping between doped infinite-layer and multilayer nickelates
exists as envisoned, some caution has to be taken concerning the details. Still in the
end, the present results for the $p=3,5$ multilayers are consistent with our previous
results on infinite-layer nickelates~\cite{lec20-1,lec20-2,lec21-1,lec21-2}: the
eventually arising superconductivity should be unconventional in a non-cuprate manner,
with the Ni-$e_g$ interplay, and especially the Ni-$d_{z^2}$ flat-band physics at
the Fermi level, as a key driving force.

\begin{acknowledgments}
The author thanks D. S. Dessau, A. Hampel, P. Hao, J. Karp, H. Li and A. J. Millis for 
helpful discussions.
Support from the European XFEL and the Center for Computational Quantum Physics
of the Flatiron Institute under the Simons Award ID 825141 is acknowledged.
The work is furthermore partially supported by the German Research Foundation
within the bilateral NSFC-DFG Project ER 463/14-1.
Computations were performed at the JUWELS Cluster of the J\"ulich
Supercomputing Centre (JSC) under project numbers hhh08 and miqs.
\end{acknowledgments}

\bibliography{bibmulti}

\end{document}